\documentstyle[12pt,psfig]{article}

\makeatletter
\@addtoreset{equation}{section}

\def\baselinestretch{1.2}
\def\href#1#2{#2}  

\newcommand{\gone}[1]{{}}
\newcommand{\norm}[1]{\raise.3ex\hbox{:} #1 \raise.3ex\hbox{:}\,}

\newcommand{\beq}{\begin{equation}}
\newcommand{\eeq}{\end{equation}}

\begin{document}

\begin{titlepage}

\begin{flushright}
NSF-ITP-00-063\\
hep-th/0008016\\
\end{flushright}
\vfil\vfil

\begin{center}

{\Large {\bf Large branes in AdS and their field theory dual}}

\vfil

\vspace{10mm}

Akikazu Hashimoto$^a$, Shinji Hirano$^a$, and N. Itzhaki$^b$\\

\vspace{10mm}

$^a$Institute for Theoretical Physics\\ University of California,
Santa Barbara, CA 93106\\
aki@itp.ucsb.edu, hirano@itp.ucsb.edu\\

\vspace{10mm}

$^b$Department of Physics\\ University of California,
Santa Barbara, CA 93106\\
sunny@physics.ucsb.edu\\

\vfil
\end{center}

\begin{abstract}
\noindent Recently it was suggested that a graviton in $AdS_5 \times
S^5$ with a large momentum along the sphere can blow up into a
spherical D-brane in $S^5$. In this paper we show that the same
graviton can also blow up into a spherical D-brane in $AdS_5$ with
exactly the same quantum numbers (angular momentum and energy). These
branes are BPS, preserving 16 of the 32 supersymmetries. We show that
there is a BPS {\it classical} solution for SYM on $S^3\times R$ with
exactly the same quantum numbers.  The solution has non-vanishing
Higgs expectation values and hence is dual to the large brane in AdS.
\end{abstract}

\end{titlepage}
\renewcommand{\baselinestretch}{1.05}  

\section{Introduction}

The holographic principle \cite{'tHooft} is a reformulation of
theories with gravity as a theory without gravity in fewer dimensions.
Such a reformulation implies a one-to-one correspondence for the
degrees of freedom and the observables between the bulk and the
boundary. This seems unlikely at first sight because these theories
live on space-times of different dimensions. The critical ingredient
that makes the holographic principle work is the presence of gravity
in the bulk theory.  It was shown, in a number of examples, that
distinct states which are indistinguishable from the boundary in the
absence of gravity become distinguishable when the effects of gravity
are taken into account \cite{Susskind}. Nonetheless, the complete
account of bulk/boundary correspondence is beyond our present
understanding of holography.

The most concrete realization of the holographic principle known to date
is the AdS/CFT correspondence \cite{Maldacena}.  The holographic
mapping of the degrees of freedom and the observables are much better
understood for this class of theories. The canonical example of this
correspondence is the duality between ${\cal N}=4$ supersymmetric
Yang-Mills theory in 3+1 dimensions (boundary) and the type IIB string
theory on $AdS_5 \times S^5$ (bulk). Under this correspondence, the
Kaluza-Klein modes of the supergravity fields on $AdS_5$ are
identified with the chiral primary operators on the SYM. This mapping
is justified in part by the fact that both sides assemble into short
supermultiplets of the superconformal algebra.  In the Hamiltonian
treatment of the SYM on $R \times S^3$, these chiral primary operators
can be associated to the physical states of the theory created by
acting with the operator on the vacuum in the infinite past. On the
AdS side, such a state corresponds to exciting the associated
Kaluza-Klein mode. The energy of such an excited state (in units of
the AdS radius) is the dimension of the chiral primary operator (see
section 3.3 of \cite{Witten}). This  provides a concrete
identification of some of the states in the boundary and in the bulk.

In a recent paper, it was suggested on the contrary that this picture
should be drastically different when the effect of angular momentum
along $S^5$ is taken into account \cite{McGreevy:2000cw}. These
authors considered Kaluza-Klein excitations carrying some angular
momentum along the $S^5$ and considered the possibility that there
exists a stable configuration of spherical branes in $S^5$ carrying
the same quantum numbers.  Although spherical branes are unstable
against shrinking due to their own tensions in the trivial vacuum,
there is an additional repulsive force due to the coupling to the
background Ramond-Ramond field in its presence.  The authors of
\cite{McGreevy:2000cw} found that there indeed exists a stable
spherical brane configuration.

On one hand, the observation of \cite{McGreevy:2000cw} is very
intriguing. The size of the spherical brane grows with angular
momentum. However, since the size of the brane can not exceed the size
of the $S^5$, there is a bound on the allowed angular momentum for the
spherical branes. This appears to offer a natural explanation for the
stringy exclusion principle \cite{Maldacena:1998uz}. However, one very
important puzzle is raised by the existence of such a spherical brane.
There appear to be {\em two} states on the supergravity side
corresponding to the state created by the chiral primary operators on
the SYM side.  This raises several questions regarding the nature of
holographic principle in AdS/CFT correspondence. Which of these states
should one associate with the chiral primary operators on the SYM
side? More importantly, what is the SYM interpretation of the states
not corresponding to the chiral primaries? One should be able to
address these questions in order to resolve the holographic
dictionary.

In this article, we will demonstrate that the situation is even more
complicated. In addition to the stable configuration of spherical
3-branes in $S^5$, there is yet another stable configuration of
spherical 3-brane in $AdS_5$ with exactly the same quantum numbers.
One must therefore find the appropriate SYM interpretation to all of
these brane configurations.

This paper is organized as follows. We will begin in section 2 by
briefly reviewing the spherical D3-branes in the $S^5$. In section 3,
we will construct the configuration of spherical D3-branes in $AdS_5$
and describe some of its properties. In section 4, we will construct a
classical solution to the equation of motion of SYM which shares much
of the properties of the spherical brane solution of section 3. We
will conclude in section 5. Some useful formulas are collected in the
appendices.

\section{Spherical Branes in $S^5$}

Let us begin by reviewing the original argument for the existence of
stable spherical brane configurations in $S^5$ \cite{McGreevy:2000cw}.
We will work with $AdS_5$ in global coordinates
\begin{eqnarray}\label{l}
ds^2 & = & {R^2 \over \cos^2\rho} \left(-d \tau^2 + d \rho^2 +
\sin^2\rho\, d \Omega_3^2\right) + R^2 d \Omega_5^2 , \nonumber \\
C_{t\Omega_3\Omega_3\Omega_3} &=& T R^4 \tan^4\rho,
\label{ads5metric}
\end{eqnarray}
where $R$ is the radius of $AdS_5$. 

Consider a graviton with angular momentum $L$ along the $S^5$.  In the
presence of the background 5-form field strength, one might expect
such a graviton to lower its own energy by ``blowing up'' into a
spherical D3-brane along the lines of Myers' mechanism described in
\cite{Myers:1999pso}.  This can not actually happen in this case
because the graviton saturates the BPS bound and its energy can not be
made any smaller.\footnote{Even in the original context of dielectric
D0-brane described in \cite{Myers:1999pso}, classical instability for
the D0-branes to blow up into a spherical D2-brane is stabilized when
the gravitational back reaction of the background RR field strength is
taken into account. We will elaborate further on this point in the
appendix A.}  At best, one can expect to find a spherical brane
carrying the same energy $E = P = L/R$ as the graviton.

\begin{figure}[t]
\centerline{\psfig{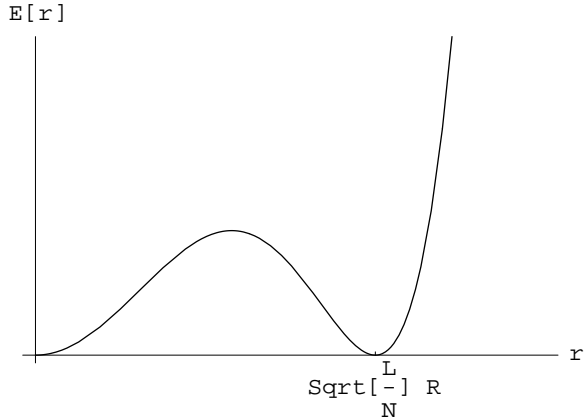}}
\caption{Energy of spherical brane as a function of its radius. The
local minima of this curve corresponds to classically stable brane
configurations.\label{figa}}
\end{figure}

Let us see that this is indeed the case.  The Lagrangian for this
system is given by
\beq
{\cal L} = -T R  \Omega_3  r^3 \sqrt{1 - (1-r^2/R^2) \omega^2} +
 \omega  N {r^4 \over R^4}.
\eeq
where 
\beq \omega = {d \phi \over d \tau}\ , \eeq
and $\phi$ is the angular parameter along the equator of $S^5$.  Due
to the rotational invariance along the equator, the angular momentum
$ L = {\partial {\cal L} / \partial\omega} $ is
conserved. Similarly, the conserved energy (in units of $1/R$) is
\beq
E = \omega L - {\cal L} = \sqrt{{N^2 r^6 \over R^6} + {(L - N
r^4/R^4)^2 \over 1-r^2/R^2}}.
\eeq
The energy $E$ as a function of $r$ is illustrated in figure
\ref{figa}.  The local minimum of $E$ at $r = \sqrt{{L \over N}} R$
corresponds to the stable configuration of spherical D3 brane of that
radius.
The fact that $\sqrt{{L \over N}} R$ must be smaller than $R$ places a
bound $L\leq N$ on the angular momentum, which was interpreted in
\cite{McGreevy:2000cw} as the manifestation of the stringy exclusion
principle \cite{ms}. Looking at the form of figure \ref{figa},
however, it is clear that there is another minimum at $r=0$. This is a
perfectly good solution to the equation of motion, at least
classically\footnote{Due to its very small size, there will be a
strong curvature correction to the DBI action.}.  Moreover, it is
clear that this minimum exists also for $L>N$.  This raises a serious
conundrum in AdS/CFT correspondence: If the minimum at $r = \sqrt{{L
\over N}}R$ is to correspond to the chiral primary operators (to match
with the stringy exclusion principle), to what does the minimum at
$r=0$ correspond? To make matters worse, we will show that there is
one more configuration of spherical D-brane in $AdS_5$ carrying the
same energy and angular momentum in the following section.

\section{Spherical branes in $AdS_5$ \label{secc}}

\begin{figure}[t]
\centerline{\psfig{file=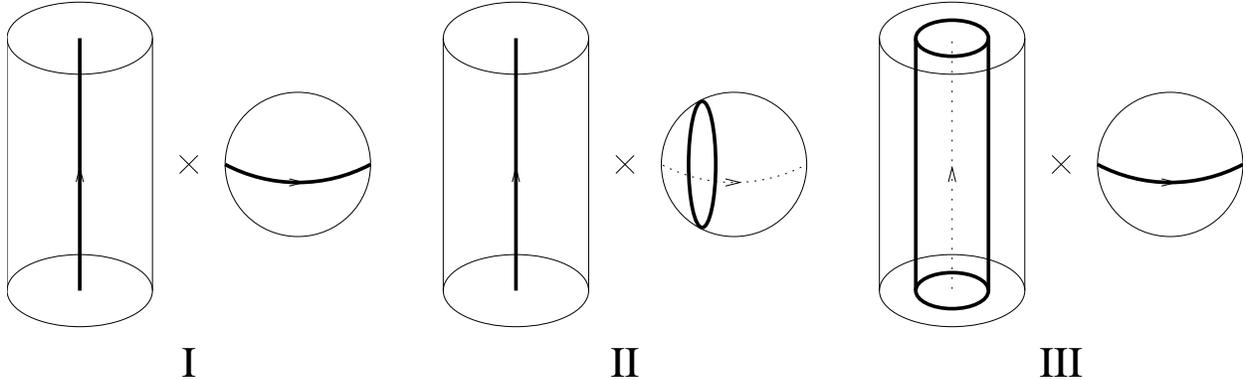,width=\hsize}}
\caption{{\bf I} collapsed spherical D3-brane of zero size, {\bf II}
spherical D3-brane embedded in $S^5$, and {\bf III} spherical D3-brane
embedded in $AdS_5$. These states are degenerate in energy and angular
momentum quantum numbers.\label{figb}}
\end{figure}

Existence of static spherical configuration of D3-branes in $AdS_5$
can be investigated along the similar lines as in the previous
section. Consider embedding a spherical D3-brane wrapping the
$\Omega_3$ of the $AdS_5$ background (\ref{ads5metric}). Just as in
the previous section, let us consider the situation where the brane is
orbiting along the equator of $S^5$ with angular velocity
$\omega$. The DBI action of such a brane configuration is
\begin{equation}
{\cal L} = - \left(T \sqrt{(-g_{tt} - \omega^2 g_{\Omega_5 \Omega_5})
g_{\Omega_3 \Omega_3}^3} - C_{t \Omega_3 \Omega_3 \Omega_3 } \right) =
-T \Omega_3 R^4 \left( \tan^3\rho \sqrt{\sec^2\rho - \omega^2} -
{\tan^4\rho} \right).
\end{equation}
Just as in the previous section, the angular momentum $L = \partial
{\cal L} / \partial \omega$ is a conserved quantity, and the canonical
energy takes the form
\begin{equation}
E = N \left( \sec \rho \sqrt{{L^2 \over N^2}+\tan^6\rho} -
{\tan^4\rho} \right) ,
\label{potential}
\end{equation}
where we have used the fact that $T \Omega_3 R^4 = N$.  This function has
essentially the same form as what is illustrated in figure
\ref{figa}. There are local minima at $\tan\rho=0$ and $\tan\rho =
\sqrt{{L \over N}} $ where $E$ takes the value (in units of $1/R$)
\begin{equation}\label{i}
E = L. \label{energy}
\end{equation}
This establishes the fact that there exists a stable configuration of
spherical D3-brane embedded in the $AdS_5$. These brane
configurations, as well as the brane configurations described in the
previous section, are illustrated in figure \ref{figb}.

\begin{figure}
\centerline{\psfig{file=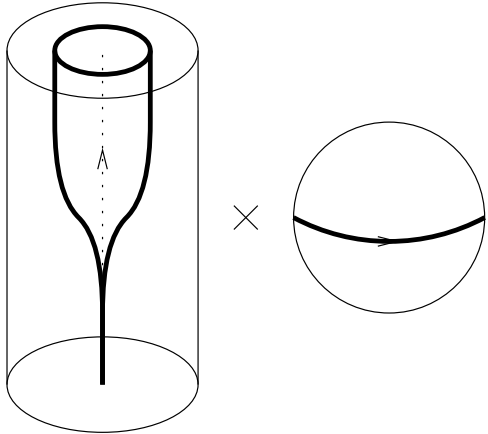}}
\caption{Instanton configuration describing the tunneling between
configurations {\bf I} and {\bf III} of figure
\ref{figb}.\label{figc}}
\end{figure}

Several comments are in order regarding the spherical brane
configuration in $AdS_5$.
\begin{itemize}

\item The spherical brane in $AdS_5$ couples electrically to the
background Ramond-Ramond field and should be thought of as a
dielectric brane. The spherical brane in $S^5$ couples magnetically
and should be thought of as a dimagnetic brane.

\item There are two solutions, one at $\tan\rho = \sqrt{L/N}$ and the
other at $\tan\rho = 0$, just as in the previous section. All of these
brane configuration preserve the same 16 of the 32 
supersymmetries of type IIB theory on $AdS_5 \times S^5$.  At first
sight this is natural for they saturate the BPS bound.  Nonetheless,
this is a very non-trivial statement since different patches of the
brane world volume are oriented in different directions.  The details
are explained in appendix B. 
 
\item There is an instanton solution describing the tunneling between
these two minima, given by
\beq
\tau = \tau_0 \pm {1 \over 2} \log \left( {\sin^2 \rho \over L/N - \tan^2
\rho } \right),  \label{tunneling}
\eeq
whose action evaluates to
\beq
S = {N \over 2} \left( {L \over N} - \log\left(1 + {L \over N} \right) \right).
\eeq
The form of this instanton solution\footnote{Similar solution
describing tunneling between spherical brane in $S^5$ and the
point-like brane also exists
\beq \tau = \tau_0 \pm {1 \over 2} \log \left({L \over N} {R^2 \over r^2}-1 \right) \label{tunneling2} \eeq
whose action evaluates to
\beq S = -{N \over 2} \left( {L \over N} + \log\left(1 - {L \over N}
\right) \right).  \eeq} is illustrated in figure
\ref{figc}.
To fully appreciate the effect of these instanton solutions, as well
as the ones on the sphere, one must take the fermion zero modes into
account. It turns out that all of these instantons are exactly $1/4$
BPS, preserving 8 of the 32 supersymmetries, as will be explained in
detail in appendix B.  The supersymmetries broken by the instantons
will give rise to fermionic zero modes which will suppress the mixing
between the two minima via the tunneling effects.

\item All of the solutions $\tan\rho = \sqrt{L/N}$, $\tan\rho = 0$, $r
= \sqrt{L/N} R$, and $r=0$ have the same energy and angular momentum
quantum numbers. 

\item Since $\tan\rho$ is not bounded, $L$ can be much larger than $N$
and hence there is no apparent connection between spherical branes in
$AdS_5$ and the stringy exclusion principle.

\item These configurations are special case of the ``large brane''
configuration discussed in related contexts in
\cite{Maldacena:1998uz,Seiberg:1999xz,Witten:1999xp,Maldacena:2000hw}. In
general, these large brane configurations are time dependent solutions
corresponding to the vacuum decay and other related phenomena.  When
the effect of both angular momentum and the Ramond-Ramond field is
taken into account, we find a novel stationary configuration of these
large branes.

\end{itemize}

The mere existence of these brane configuration raises an important
question: How does one distinguish between these states from the
viewpoint of the boundary theory? Unfortunately, we are unable to
offer a complete resolution to this problem.  One very concrete and
interesting observation that we discuss in the next section is that
the spherical branes in $AdS_5$ (as opposed to the spherical branes in
$S^5$ and the point-like brane) turns out to have a concrete
interpretation as a {\it classical} solution from the field theory
point of view.

\section{Spherical branes in $AdS_5$ as classical solutions of SYM}

In this section, we will describe a solution to the classical equation
of motion of the SYM which is dual of the spherical
branes in AdS.

Configuration of spherical branes in $AdS_5$ (illustrated in figure
\ref{figb}.{\bf III}) is such that the flux of RR 5-form in the
interior of the spherical D3-brane is less by one unit compared to the
exterior.  In light of the UV/IR relation of the AdS/CFT
correspondence, this suggests that the gauge symmetry is broken from
$SU(N)$ to $SU(N-1)\times U(1)$ at low energies.  Therefore we should
look for a classical configuration involving Higgs expectation values.

Since the D3-branes do not act as a source for the dilaton and the
axion, the supergravity back reaction of the spherical D3-branes is
trivial in the dilaton/axion sector. Trivial dilaton/axion background
corresponds to trivial $F^2$ and $F \tilde F$ expectation values. The
field theory counterpart of the spherical brane is therefore not likely 
to involve the gauge fields.
Furthermore, the fact that the energy (\ref{i}) of the solution we are after
does not depend on the coupling constant suggests that the commutator
term in the action of the SYM should not play any role. We are
therefore left with the Abelian part of the action of the six scalar
fields $\phi_i$, $i=1 \ldots 6$. 

Theories on $S^3\times R$ contain an additional term in the action
coming from the positive curvature of $S^3$.  In $n$ dimensions this
term is fixed by the conformal invariance to be
\begin{equation}
S = -{1 \over 2g_{YM}^2} \int d^n x \left( (\partial \phi_1)^2 +
(\partial \phi_2)^2 + {(n-2) \over 4(n-1)} \tilde{R}
(\phi_1^2+\phi_2^2)\right) \ , \label{action}
\end{equation}
where $\phi_1$ and $\phi_2$ are the two scalars we focus on and
$\tilde{R}$ is the Ricci curvature which is related to the radius $R$
of $AdS_{n+1}$ by
\begin{equation}
\tilde{R} =  {(n-1) (n-2) \over R^2}.
\end{equation}
Setting $n=4$, the action becomes
\begin{equation}
S = {R^{3} \Omega_{3} \over 2 g_{YM}^2} \int dt \  \left(  \dot{\phi}_1^2 + 
\dot{\phi}_2^2 - {1 \over
R^2}(\phi_1^2+\phi_2^2) \right).
\end{equation}
Reparameterizing the fields according to 
\begin{equation}
\phi_1 = \sqrt{g_{YM}^2 N \over R^2 \Omega_3 } \eta \cos\theta , \qquad
\phi_2 = \sqrt{g_{YM}^2 N \over R^2 \Omega_3} \eta \sin\theta,
\end{equation}
gives
\begin{equation}
S = {N R \over 2 } \int dt\ \left(  \dot \eta^2 +
\eta^2 \dot \theta^2 -{\eta^2 \over R^2} \right).
\end{equation}
Now, consider the ansatz
\begin{equation}
\eta = {\rm const.}, \qquad \theta = \omega t \label{ansatz}.
\end{equation}
The angular momentum $L = dS/d\omega$ is conserved, and the conserved
energy, in units of $1/R$ (see eq.(\ref{l})), is
\begin{equation}
E = L \omega - {\cal L} = \left({ L^2 \over 2 N \eta^2} + { N \eta^2 \over 2}
\right) \label{YMenergy}
\end{equation}
which is minimized at
\begin{equation}
\eta = \sqrt{{L \over N}} . \label{rhomin}
\end{equation}
This constitutes a solution to the equation of motion of the field
theory (\ref{action}). The energy associated with this solution is 
\begin{equation}
E =  L. \label{YMminenergy}
\end{equation}

To properly account for the $SU(N)$ field content of the SYM, simply
parameterize $\phi_1$ and $\phi_2$ according to
\begin{equation}
\phi_1 =\sqrt{ g_{YM}^2 N\over R^2 \Omega_3 } \hat\eta \cos\theta,
\qquad \phi_2 = \sqrt{ g_{YM}^2 N\over R^2 \Omega_3 } \hat\eta
\sin\theta\ ,
\end{equation}
where $\hat{\eta}$ is a traceless diagonal $N \times N$ matrix
\begin{equation}
\hat\eta = 
\sqrt{{N-1 \over N}}
\left(\begin{array} {cccc}
\eta   &	              &        & \\
       & -{\eta  \over N-1}   &        & \\
       &		      & \ddots & \\
       &		      &        & -{\eta  \over N-1} 
\end{array}\right)\ .
\end{equation}
To leading order in $1/N$, all but the first diagonal element can be
ignored and the analysis reduces to treating $\phi_{1,2}$ as an
ordinary scalar field. The subleading $1/N$ correction can be thought
of as the back reaction of the spherical brane to the background
geometry. Taking the full matrix structure of $\phi_{1,2}$ into
account does not affect (\ref{YMenergy}-\ref{YMminenergy}) for they
commute.

Let us make some comments regarding this solution
\begin{itemize}

\item The energy of the classical solution (\ref{YMminenergy}) is
precisely the energy of the spherical brane in $AdS_5$ found in
equation (\ref{energy}).

\item The magnitude of the scalar expectation value (\ref{rhomin}) is
the same as the SUGRA result if one uses the original UV/IR relation
of Maldacena \cite{Maldacena} and not the ones of
\cite{Susskind:1998dq}. This is expected for we are dealing with Higgs
expectation values and not gravitational waves as the probes in the bulk.

\item The classical solution is invariant under half of the
supersymmetries. This can be verified easily by acting on the solution
with the supersymmetry transformation rules given in
\cite{Nicolai:1988ek}. (Strictly speaking, we have only checked that
the solution is invariant with respect to 8 out of 16 Poincare 
supersymmetries.)

\end{itemize}

\begin{figure}
\centerline{\psfig{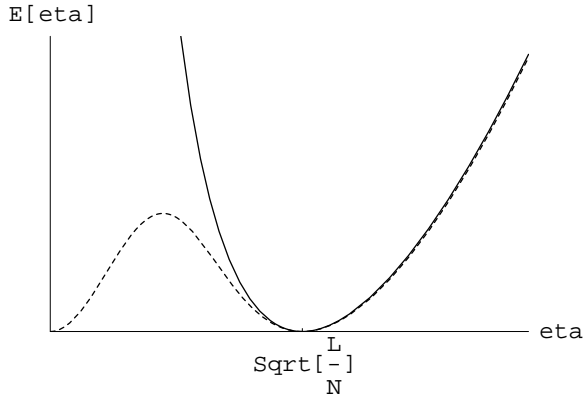}}
\caption{Solid line is energy as a function of $\eta$ of the classical
solution (\ref{YMenergy}) applicable for small $\lambda$. The dotted
line is the same function for brane probe action (\ref{potential2})
applicable for large $\lambda$. \label{figd}}
\end{figure}

The fact that the classical solution of the SYM shares many properties
in common with the spherical brane configuration in $AdS_5$ is a good
indication that the former is the field theory realization of the
latter. There are some subtle differences, however. The potential
(\ref{YMenergy}) is the field theory counterpart of
(\ref{potential}). To be more precise, (\ref{potential}) is the
effective action for the spontaneously broken $U(1)$ at large
$\lambda$ after integrating out the massive W-bosons.  Equation
(\ref{YMenergy}) can simply be thought of as the small $\lambda$ limit
of the same quantity.  To facilitate the comparison, let us re-express
(\ref{potential}) in terms of $\eta = \tan \rho$
\beq E = N \left( \sqrt{1 + \eta^2} \sqrt{{L^2 \over N^2} + \eta^6} -
\eta^4 \right) \ .
\label{potential2}
\eeq
Potentials (\ref{YMenergy}) and (\ref{potential2}) differ from each
other in one very important sense. (See figure \ref{figd} for an
illustration.) The potential at strong coupling (\ref{potential}) has
two minima, one at $\eta =0$ and the other at $\eta =\sqrt{L/N}$. At
small coupling, (\ref{YMenergy}) has only one minima, at $\eta =
\sqrt{L/N}$.

What happened?  What we have found is an argument based on duality
that the minima at $\eta=0$ is lifted by $1/\lambda$ corrections.
When $\lambda \ll 1$, semi-classical description of the SYM becomes
reliable, but the configuration at $\eta=0$ simply does not exist as a
solution of the classical equation of motion.  It would be very
interesting to understand the status of $\eta=0$ solution when the
quantum effects on the SYM side is taken into account. Studying the
quantum correction to (\ref{YMenergy}) perturbatively should teach us
a lot about this issue. 

Unlike the solution at $\eta=0$, the solution at $\eta=\sqrt{L/N}$ is
a robust result.  This can be seen in the following way.  For large
values of $L/N$, the spherical brane will grow to have size much
greater than the radius of $AdS_5$.  In \cite{Seiberg:1999xz}, Seiberg
and Witten showed that the DBI+CS action of the $n$-brane in
$AdS_{n+1}$ has the following form for $n>2$ near the boundary of the
AdS (see eq.(3.17) of that paper)
\begin{equation}\label{m}
S\sim \int \sqrt{g} \left( (\partial \phi)^2 +\frac{n-2}{4(n-1)}\phi^2
\tilde{R} +{\cal O}(\phi^{\frac{2(n-4)}{n-2}})\right)\ . \label{sw}
\end{equation}
The form of this action is dictated by the fact that the extension of
the metric on the boundary of AdS to the bulk is unique in the
neighborhood of the boundary \cite{gl}.  The leading term in large
$\phi$ of (\ref{sw}) exactly matches the field theory action
(\ref{action}).

\section{Conclusions}

The main goal of this paper is to point out an important subtlety in
our current understanding of holography and AdS/CFT correspondence.
In AdS/CFT, there is a natural one-to-one correspondence between the
chiral primary operators of the boundary theory and the Kaluza-Klein
excitations on the bulk. However, there exists a configuration of
spherical D-branes embedded in the $S^5$ {\em in addition} to the
Kaluza-Klein excitations, carrying the same quantum numbers as was
demonstrated in \cite{McGreevy:2000cw}.  In this article, we
demonstrated that there is {\em yet another} configuration of
spherical D-branes, embedded in the $AdS_5$, again with the same
quantum numbers. The full understanding of holographic principle will
require that one understands how each of these spherical branes are
realized on the field theory side.

The spherical branes in AdS are much like the long strings
\cite{Maldacena:1998uz,Seiberg:1999xz,Maldacena:2000hw} in $AdS_3
\times S^3$ (See also appendix C). The long strings live on the
boundary of $AdS_3$, and gives rise to new class of operators of the
CFT. The spherical branes in higher AdS appears to play a slightly
different role. These branes do not live at the boundary but at some
definite radius in the bulk.  Unlike the long strings, these branes
are completely degenerate in angular momentum and energy with the
Kaluza-Klein excitations.

We have not resolved the problem of identifying and distinguishing all
of the brane configuration from the field theory side.  To partially
address this problem, we described a classical solution to the
equation of motion of the SYM which shares many of the properties of
the spherical branes in $AdS_5$. The collapsed brane configurations
and the spherical branes in $S^5$ do not appear to correspond to a
classical solution in a similar manner. Does this also imply that the
$r=0$ solution of \cite{McGreevy:2000cw} is also lifted? This depends
on whether the $r=0$ solution and the $\rho=0$ solution can be
identified as the same physical state.  This is a tricky question
because there is a large degeneracy of states that look like figure
\ref{figb}.{\bf I} especially when multi-particle states are taken
into account. More detailed understanding of the holographic map is
needed to resolve this issue. Even if the $\rho=0$ state and the $r=0$
state turns out not to be the same physical state, the fact that the
$\rho=0$ solution was lifted by $1/\lambda$ correction is a strong
indication that the $r=0$ solution is also lifted.

From the point of view of semi-classical SYM, BPS classical solution
is a coherent state of many quanta of chiral excitations. If the
identification of spherical branes in $S^5$ with the states created by
the chiral primary operators turns out to be correct, this suggests
that the spherical brane in $AdS_5$ is a coherent state of spherical
branes in $S^5$. It would be very interesting to understand this point
better.

In order to proceed further, it appears to be necessary to properly
address either the quantum correction of the SYM side or the curvature
correction of the supergravity side. This is clearly a non-trivial
challenge.  The spherical brane configurations in supergravity do
exist, and their existence is a prediction about strongly coupled
gauge theory via the AdS/CFT correspondence. Learning to resolve these
redundancies should teach us a lot about the dynamical aspects of
quantum field theories, as well as the holographic principle.

\section*{Acknowledgments}

We would like to thank O.~Aharony, M.~Berkooz, J.~David, D.~Gross,
G.~Horowitz, J.~Maldacena, J.~McGreevy, S.~Minwalla and J.~Polchinski
for discussions.  The work of A.H.\ is supported in part by the NSF
Grant No.~PHY94-07194. The work of S.H.\ is supported in part by the
Japan Society for the Promotion of Science. The work of N.I.\ is
supported in part by the NSF grant No.~PHY97-22022.

\section*{Appendix A: Classical Stability and the Gravitational Back Reaction}
        {\setcounter{section}{1} \gdef\thesection{\Alph{section}}}
	 {\setcounter{equation}{0}}

The general feature that branes of spherical geometry can exist stably
in a background of some anti-symmetric tensor field strength is quite
similar to the mechanism of dielectric branes discussed by Myers
\cite{Myers:1999pso}, but there is one critical difference.  In AdS,
the $r=0$ solution is classically stable. In Myers' analysis, $r=0$
solution is classically unstable. In this appendix, we will explain
that even in Myers' example, $r=0$ solution is classically stable when
the effect of gravitational back reaction of the stress-energy of the
background Ramond-Ramond field strength is taken into account.

\begin{figure}
\centerline{\psfig{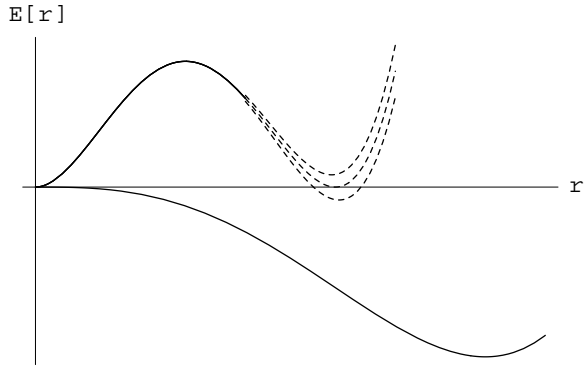}}
\caption{Energy of spherical brane in the background Ramond-Ramond
field neglecting the back reaction (\ref{myers87}) and including the
back reaction (\ref{a7}). These curves are only valid locally near
$r=0$. When the back reaction is taken into account, $r=0$ is
classically stable.  However, it does not determine if this point is a
global minimum or not.\label{figg}}
\end{figure}

Myers' analysis assumes a flat space-time in a background of constant
RR 4-form field strength, giving rise to a potential of the form (see
equation (87) of \cite{Myers:1999pso}.)
\begin{equation}
E(r) = T_2 \left(\sqrt{{\alpha'^2 N^2 \over 4} + r^4} - F r^3\right)
\approx T_2 \left( {N \alpha' \over 2} - F r^3 + {r^4 \over N
\alpha'}\right) \ . \label{myers87}
\end{equation}
The relevant energetic consideration comes from the $r^3$ term which
has a negative coefficient, and the $r^4$ term which has a positive
coefficient. Since the leading small $r$ effect is negative, there is
a classical instability (see figure \ref{figg}).

The essential difference between Myers' flat space analysis and our
analysis in AdS is that in AdS, there is a term in $E(r)$ which grows
quadratically, thereby dominating over the cubic term which is the
leading contribution in Myers' analysis. Closer examination reveals
that the quadratic term arises from the $r$ dependence of $g_{00}$
which enters into the Nambu-Born-Infeld action. $g_{00}$ has
non-trivial $r$ dependence because the space-time is curved in
response to the stress energy generated by the cosmological constant
in the AdS space.

However, there is also stress energy associated with the field
strength in the system considered by Myers. The stress energy due to
the background field strength can certainly give rise to a non-trivial
$r$ dependence in the background metric. This is the effect of
gravitational back reaction of the anti-symmetric form field strength
background which was ignored in Myers' analysis. However, if the
effect of such a back reaction enters at quadratic order, it can
drastically alter the conclusion regarding the classical instability
in the small $r$ region.

We claim that this is indeed the case.  A complete analysis for that
issue in full generality is beyond the scope of this paper.  Instead
we shall demonstrate that point by considering an example which is
sufficiently generic.  We consider a version of Myers' mechanism where
a D1-brane is blown up into a spherical D3-brane in the presence of RR
5-form field strength.  The reason that this should be considered
sufficiently generic is the fact that in string theory, RR 5-form
background can only be created using a source that is available in the
theory.  The object which acts as a source for RR 5-form field
strength is the D3-brane. D3-branes are especially convenient because
they do not act as a source for the dilaton.\footnote{This is not the
most general possibility. For example, RR waves with no D-branes
sources are excluded from this class of backgrounds.}

Consider taking a large number of D3-branes oriented along the 0123
directions, distributed arbitrarily along the 456789 directions.  Our
ability to construct consistent backgrounds will be parameterized by
the degree of freedom in distributing the  D3-branes. The general form
of backgrounds that can be generated this way takes the form \cite{hs}
\begin{equation}
ds^2 = f^{-1/2} (-dt^2+dx_1^2 + dx_2^2 + dx_3^2) + f^{1/2} (dx_4^2 +
dx_5^2+dx_6^2+dx_7^2 + dx_8^2+dx_9^2)
\end{equation}
\begin{equation}
F_{0123i} = \partial_i f^{-1}
\end{equation}
where $f(x_4,x_5,x_6,x_7,x_8,x_9)$ is an arbitrary  harmonic
function on the 456789-plane.

At some fixed point in the 456789 plane, the invariant field strength
squared is given by
\begin{equation}
F^2 = {(\partial_i f)^2 \over f^{5/2}} \ .
\end{equation}
Using only the fact that $f$ is a harmonic function ($\nabla^2 f =
0$), one finds that at the neighborhood of that point
\begin{equation}
R_{ij} = -F^2 g_{ij}, \qquad R_{ab} = F^2 g_{ab}
\end{equation}
where $i$ and $j$ runs over $0123$ and the direction of the gradient
$\vec\nabla f$, whereas $a$ and $b$ runs over the rest of the
directions. Therefore, at any point in the 456789 plane, the local
geometry, to quadratic order in geodesic distance, is an $AdS_5 \times
S^5$ with the radius of the order $R^2 = 1/F^2$. This in turn implies
that there will always be a quadratic correction to the $g_{00}$
component of the metric in the locally inertial frame
\begin{equation}
g_{00} \approx  -(1 + F^2 r^2 + \ldots)\ .
\end{equation}
This term will always give rise to a quadratically rising potential in
$E(r)$ 
\begin{equation}
E(r) = T_2 \left( \sqrt{-g_{00} \left({\alpha'^2 N^2 \over 4} + r^4\right)} -   F r^3\right) \approx T_2 \left({N \alpha' \over 2}  + N \alpha' F^2 r^2 - F r^3 + \ldots \right) \label{a7}
\end{equation}
which dominates over the $r^3$ term at small $r$. Because this is the
leading effect at small $r$, it is inconsistent to ignore the effect
of gravitational back reaction. When this effect is properly taken
into account, there will never be a classical instability at $r=0$ for
a freely falling D1-brane. It should be emphasized however that this
discussion is valid only locally.  In general it might be that the
minimum at $r=0$ is only a local minimum and not a global minimum.

\section*{Appendix B: Supersymmetry condition for branes in $AdS_5 \times S^5$}
        {\setcounter{section}{2} \gdef\thesection{\Alph{section}}}
	 {\setcounter{equation}{0}}

In this appendix, we will analyze the supersymmetric properties of the
spherical branes in $AdS_5$ and $S^5$, as well as the instanton
solutions describing the tunneling between the degenerate vacua. The
strategy is to simply apply the supersymmetry condition
\cite{Cederwall:1997ri,Bergshoeff:1997kr} locally and to count the
supersymmetries that are left unbroken globally. Similar strategies
have been applied to study the global supersymmetries of baryonic
configurations in $AdS_5 \times S^5$
\cite{Imamura:1998gk,Gomis:1999xs}. The main complication arises from
the fact that both the background and the brane are curved. Following
\cite{Imamura:1998gk}, let us introduce
\beq \epsilon^\pm(x) = \epsilon_L(x) \pm i \epsilon_R(x) \eeq
where $\epsilon_{L,R}(x)$ are the left and right handed Majorana-Weyl
Killing spinors of type IIB supergravity on $AdS_5 \times S^5$, with
positive spacetime chirality $\Gamma_{11}=+1$. The local supersymmetry
condition for D3-branes under consideration takes the form  
\beq \Omega^{ijkl}(x) \Gamma_{ijkl} \epsilon^{\pm}(x) =  \mp i
\epsilon^{\pm}(x) 
\eeq 
where $\Omega^{ijkl}(x)$ is proportional to the volume element of the
D3-brane. It is 
convenient to write the covariantly constant spinors $\epsilon^{\pm}(x)$
in the form
\beq \epsilon^{\pm}(x)  = S^{\pm}(x) \epsilon_0^{\pm} \eeq
so that the local supersymmetry condition reads
\beq \Omega^{ijkl}(x) \Gamma_{ijkl} S^{\pm}(x) \epsilon_0^{\pm} = \mp
i S^{\pm}(x) \epsilon_0^{\pm} \ . \label{localsusy} \eeq
The number of independent spinors $\epsilon_0^{\pm}$ satisfying the
condition (\ref{localsusy}) for all $x$ is the number of unbroken
supersymmetries of the brane configuration. It should be emphasized
that the condition on $\epsilon_0^{\pm}$ at different values of $x$ is
an overcomplete set, and that there exist any $\epsilon_0^{\pm}$ at
all that satisfies this requirement is highly non-trivial. 

This non-trivial condition is satisfied by the brane configurations
illustrated in figure \ref{figb}. The condition on $\epsilon_0^{\pm}$
simplifies to
\beq (1 - \Gamma_\tau \Gamma_\phi) \epsilon_0^{\pm} = 0,
\label{susycond}\eeq
where $\Gamma_{\tau}$ and $\Gamma_\phi$ are the $\Gamma$ matrices 
associated with the time direction and the direction of the orbit of
the branes in $S^5$ respectively. This condition is exactly the same
as that for massless particles in ten dimensions, and the same
condition applies to all of the brane configurations illustrated in
figure \ref{figb}. In other words, these branes are indistinguishable
at the level of supersymmetries, and they all belong to the same
supermultiplet as that of the supergraviton.

Instanton solutions describing the tunneling between the spherical and
the point like branes also preserve some fraction of
supersymmetries. In addition to (\ref{susycond}), the constraint on
supersymmetires imposed by the instanton solution takes the form
\beq \Gamma_{r\phi_1\phi_2\phi_3}\epsilon_0^{\pm}=\epsilon_0^{\pm},
\label{susyinst}
\eeq
where $\Gamma_r$ and $\Gamma_{\phi_1\phi_2\phi_3}$ are the $\Gamma$
matrices associated with the radial and the three-sphere
directions. Therefore, these instantons preserve one quarter of the
supersymmetries.

In the remainder of this appendix, we will summarize the argument
leading to the conclusion that the conditions for supersymmetry are 
given by (\ref{susycond}) and (\ref{susyinst}). To proceed, we need an
explicit expression for the $\Omega^{ijkl}$ and the $S^{\pm}(x)$. Let us
begin by choosing an explicit coordinates for $AdS_5 \times S^5$.  We
will continue to use the metric (\ref{ads5metric}) and parameterize
the 3-sphere and the 5-sphere according to
\begin{eqnarray}
d\Omega_3^2 & =& d \theta_1^2 + \sin^2 \theta_1 d \theta_2^2 + \sin^2 \theta_1 \sin^2 \theta_2 d \theta_3^2 \\
d\Omega_5^2 &=& d \theta_{\bar{5}}^2
+ \sin^2 \theta_{\bar{5}} d \theta_{\bar{4}}^2
+ \sin^2 \theta_{\bar{5}}\sin^2 \theta_{\bar{4}} d \theta_{\bar{3}}^2 
+ \sin^2 \theta_{\bar{5}}\sin^2 \theta_{\bar{4}} \sin^2\theta_{\bar{3}} d \theta_{\bar{2}}^2 \nonumber \\
&& \qquad + \sin^2 \theta_{\bar{5}}\sin^2 \theta_{\bar{4}} \sin^2 \theta_{\bar{3}} \sin^2 \theta_{\bar{2}} d \theta_{\bar{1}}^2  \ .
\end{eqnarray}
We will generally use barred indecies to refer to $S^5$ coordinates
and unbarred indecies to refer to $AdS_5$ coordinates.

We will follow the $\Gamma$ matrix conventions of \cite{Lu:1998nu}:
\beq \Gamma_m = \sigma_2 \otimes \gamma_m^{AdS} \otimes 1_4, \qquad
\Gamma_{\bar{m}} = \sigma_1 \otimes 1_{4} \otimes \gamma_{\bar{m}}^S,
\qquad \gamma = \Gamma_{\bar{1}\bar{2}\bar{3}\bar{4}\bar{5}} =
\sigma_1 \otimes 1_4 \otimes 1_4, \eeq
where $m=0,1,2,3,4$ and
$\bar{m}=\bar{1},\bar{2},\bar{3},\bar{4},\bar{5}$.  An explicit form
of the Killing spinors in $AdS_5\times S^5$ can be obtained by
combining the results of \cite{Lu:1998nu} and
\cite{Breitenlohner:1982jf}
\beq S^{\pm}(x) =\sqrt{\sec\rho}\left(\cos\frac{\rho}{2}\pm
i\hat{x}^{\alpha}\gamma\Gamma_{\alpha}\sin\frac{\rho}{2}\right)e^{\pm
i{\tau \over 2}\gamma\Gamma_0}e^{\pm{i \over
2}\theta_{\bar{5}}\gamma\Gamma_{\bar{5}}}\prod_{m=4}^1 \left(e^{-{1
\over 2}\theta_{\bar{m}}\Gamma_{\bar{m},\bar{m}+\bar{1}}}\right)\ ,
\eeq
where $\hat{x}_i$ are defined following  \cite{Breitenlohner:1982jf}
\begin{eqnarray}
\hat{x}_1 & = & \sin \theta_1 \sin \theta_2 \sin \theta_3 \nonumber \\
\hat{x}_2 & = & \sin \theta_1 \sin \theta_2 \cos \theta_3 \nonumber \\
\hat{x}_3 & = & \sin \theta_1 \cos \theta_2 \nonumber \\
\hat{x}_4 & = & \cos \theta_1   \ .
\end{eqnarray}

\subsection*{Appendix B.1: Supersymmetry of the spherical branes} 

We are now ready to analyze the unbroken supersymmetries of the
spherical D3-branes in $AdS_5$. Let us take $\theta_{\bar{5}}$ to be
the direction of orbit along the $S^5$, more specifically the other
angles $(\theta_{\bar{4}},\theta_{\bar{3}},\theta_{\bar{2}})$ set to
be zero, and $\theta_{\bar{1}}$ to be $0$ or $\pi$, to cover the orbit
globally by two patches. Then the volume form of the spherical
D3-brane takes the form of
\beq  \Omega = {R^4(\sec^2\rho d\tau +\omega
d\theta_{\bar{5}} )\over \sqrt{\sec^2 \rho - \omega^2}}\wedge d\theta_3\wedge d\theta_2\wedge
d\theta_1. \eeq
Thus $\Omega^{ijkl} \Gamma_{ijkl}$ is given by
\beq
\Omega^{ijkl}\Gamma_{ijkl} = 
\frac{1}{\sqrt{\sec^2\rho-\omega ^2}}\left(\sec\rho\Gamma_0+\omega \Gamma_{\bar{5}}\right)\left(-\hat{x}_1\Gamma_{234}+\hat{x}_2\Gamma_{134}-\hat{x}_3\Gamma_{124}+\hat{x}_4\Gamma_{123}\right).
\eeq

Now for the point-like solution at $\rho=0$, $\omega = 1$ (figure
\ref{figb}.{\bf I}), and on the $\theta_{\bar 1}=0$ branch, it is easy
to show that (\ref{localsusy}) implies
\beq 
(1 - \Gamma_0 \Gamma_{\bar{5}})\epsilon_0^{\pm} = 0.
\label{answer}
\eeq
On the $\theta_{\bar{1}} = \pi$ branch, $\omega = -1$ since the
orientation of the orbit is reversed.

For the finite size solution at $\tan\rho = \sqrt{L/N}$, $\omega = 1$
(figure \ref{figb}.{\bf III}), after straightforward manipulation of
$\Gamma$ matrices, (\ref{localsusy}) also reduces to
(\ref{answer}). The same remark on the two branches of orbit is
applicable in this case as well. Note that the condition $\omega=1$
(or $\omega=-1$) is actually necessary for preserving the global
supersymmetries.

The unbroken supersymmetry of spherical branes in $S^5$ (figure
\ref{figb}.{\bf II}) can be analyzed in a similar manner. We will
again take $\theta_{\bar{5}}$ to be the direction of orbit of the
center of mass. Then the volume form takes the form of
\beq \Omega = {R^4(d\tau +\omega
(1-r^2/R^2)d\phi) \over \sqrt{1 - (1-r^2/R^2) \omega^2}}\wedge d\theta_{\bar{3}}\wedge
d\theta_{\bar{2}}\wedge d\theta_{\bar{1}}, 
\eeq
where
$\tan\phi=\tan\theta_{\bar{5}}\cos\theta_{\bar{4}}$, 
$r=R\sin\theta_{\bar{5}}\sin\theta_{\bar{4}}$, and the brane is
sitting at the origin $\rho=0$ in $AdS_5$. Thus we have
\beq \Omega^{ijkl} \Gamma_{ijkl} = 
\frac{1}{\sqrt{1-(1-r^2/R^2)\omega^2}}\left\{\Gamma_0+\omega
\left(\cos\theta_{\bar{4}}\Gamma_{\bar{5}}-{r \over
R\tan\theta_{\bar{5}}}
\Gamma_{\bar{4}}\right)\right\}\Gamma_{\bar{3}\bar{2}\bar{1}}.
\eeq
After some manipulation, once again (\ref{localsusy}) simplifies to
\beq
(1 - \Gamma_0\Gamma_{\bar{5}})\epsilon_0^{\pm}=0.
\eeq

\subsection*{Appendix B.2: Supersymmetry of the instantons}

Now we proceed to the analysis of unbroken supersymmetries of
instantons on the spherical D3-branes in $AdS_5$. The analysis goes
through in much the same way as in the previous cases. The only
difference comes from the time-dependence on the radial direction
which will be reflected in the time-like direction of the worldvolume
of the spherical D3-branes. As a result $\Omega^{ijkl}\Gamma_{ijkl}$
takes the form of
\beq
\Omega^{ijkl}\Gamma_{ijkl}=
\frac{1}{\sqrt{\sec^2\rho-\dot{\rho}^2\sec^2\rho-\omega^2}}
\left(\sec\rho\Gamma_0+\dot{\rho}\sec\rho\hat{x}^{\alpha}\Gamma_{\alpha}
+\omega\Gamma_{\bar{5}}\right)
\left(-\hat{x}^{\beta}\Gamma_{\beta}\Gamma_{1234}\right).
\eeq
After a little computation, one finds the global supersymmetry
conditions to be
\begin{eqnarray}
\left(1-\Gamma_0\Gamma_{\bar{5}}\right)\epsilon_0^{\pm}&=&0,
\label{adsgrav}\\
\Gamma_{1234}\epsilon_0^{\pm}&=&\epsilon_0^{\pm},
\label{adsquarter}\\
\omega \mp i\dot{\rho}\tan\rho -1&=&0.
\label{adsinst}
\end{eqnarray}
Using the relation between $\omega$ and $L$, it is easy to show that
(\ref{adsinst}) is the instanton equation (in Euclidean time)
\beq
\mp\dot{\rho}=\tan\rho\frac{\tan^2\rho-L/N}{\tan^4\rho+L/N}
\eeq
whose solution is (\ref{tunneling}).

Similarly on the spherical D3-branes in $S^5$,
$\Omega^{ijkl}\Gamma_{ijkl}$ is given by
\begin{eqnarray}
\Omega^{ijkl}\Gamma_{ijkl}&=&
\frac{1}{\sqrt{1-{(\dot{r}/R)^2 \over
1-(r/R)^2}-(1-(r/R)^2)\omega^2}}
\left\{\Gamma_0 +{\dot{r}/R \over 1-(r/R)^2}
\left(\cos\theta_{\bar{5}}\sin\theta_{\bar{4}}\Gamma_{\bar{5}}
+\cos\theta_{\bar{4}}\Gamma_{\bar{4}}\right)\right.\nonumber\\
&&\left.\qquad\qquad\qquad\qquad\qquad\qquad\qquad
+\omega\left(\cos\theta_{\bar{4}}\Gamma_{\bar{5}}
-{r \over R\tan\theta_{\bar{5}}}\Gamma_{\bar{4}}\right)\right\}
\Gamma_{\bar{3}\bar{2}\bar{1}}.
\end{eqnarray}
This time, the condition for preservation of global supersymmetry
turns out to be
\begin{eqnarray}
\left(1-\Gamma_0\Gamma_{\bar{5}}\right)\epsilon_0^{\pm}&=&0,
\label{s5graviton}\\
\Gamma_{\bar{1}\bar{2}\bar{3}\bar{4}}\epsilon_0^{\pm}&=&\epsilon_0^{\pm},
\label{s5quarter}\\
\omega \pm i{r \over R}{\dot{r}/R \over 1-(r/R)^2}-1&=&0.
\label{s5inst}
\end{eqnarray}
Once again one can easily verify that the last condition
(\ref{s5inst}) is precisely the instanton equation
\beq
\pm {\dot{r} \over R} = {r \over R}\left\{{N \over L}
\left({r \over R}\right)^2-1\right\}
\eeq
whose solution is (\ref{tunneling2}).

\section*{Appendix C: Generalizations to other AdS}
        {\setcounter{section}{3} \gdef\thesection{\Alph{section}}}
	 {\setcounter{equation}{0}}

In this article, we concentrated mainly on spherical branes in
$AdS_5\times S^5$. This can be generalized immediately to $AdS_7
\times S^4$ and $AdS_4 \times S^7$. Following the argument presented
in section \ref{secc}, one obtains an expression for the energy as a
function of the angular momentum $L$ and the radius $\rho$
\begin{itemize}
\item M2 in $AdS_4 \times S^7$
\beq E(\rho,L) = N \left( \sec \rho \sqrt{{L^2 \over 4 N^2} + \tan^4 \rho} - \tan^3\rho\right) \eeq

\item M5 in $AdS_7 \times S^7$
\beq E(\rho,L) = N \left( \sec \rho \sqrt{{4 L^2 \over  N^2} + \tan^{10} \rho} - \tan^6\rho\right) \ .\eeq
\end{itemize}
These potentials essentially behave like (\ref{potential}) illustrated
in figure \ref{figa}. These are the electric counterparts to the
spherical magnetic dipoles in $AdS_4\times S^7$ and $AdS_7 \times S^4$
described in \cite{McGreevy:2000cw}.

\begin{figure}
\centerline{\psfig{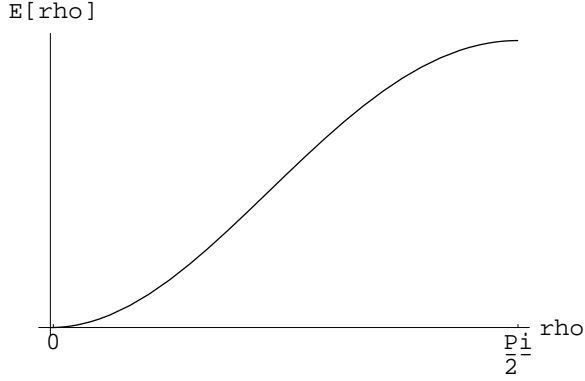}}
\caption{$E(\rho,L)$ for spherical D1-brane in $AdS_3 \times S^3$.\label{fige}}
\end{figure}

The case of $AdS_3 \times S^3$ is somewhat different. Consider a
D1-string probe in the $AdS_3 \times S^3 \times T^4$ background with
$Q_1$ and $Q_5$ units of electric and magnetic Ramond-Ramond 3-form
fluxes, respectively. The potential energy then takes the form
\beq E(\rho,L) = Q_5 \left( \sec \rho \sqrt{\left({L^2 \over Q_5^2 }\right) + \tan^2 \rho} - \tan^2\rho \right) \ . \eeq
The potential has one global minimum at $\rho=0$. There is an unstable
stationary point at $\rho = \pi/2$ (see figure \ref{fige}). This is
precisely the long string of \cite{Maldacena:2000hw}. At the special
value of angular momentum $L = Q_5$, the potential becomes completely
flat and the long and the short strings become degenerate
\cite{Maldacena:2000hw}. The fact that this happens at $L = Q_5$
rather than $L = Q_1 Q_5$ indicates that this effect is unrelated to
the stringy exclusion principle of $AdS_3 \times S^3.$

Spherical D-strings in $S^3$ can also be analyzed in similar manner.
The potential is found to take the form
\beq E = \sqrt{{Q_5^2 r^2 \over R^4} + {(L^2 - Q_5 r^2/R^2)^2 \over R^2-r^2}}
\eeq
which for $L = Q_5$ becomes flat and degenerate.

\section*{Note Added}

While this paper was in preparation, we learned that similar results
are being considered by Grisaru, Myers, and Tafjord \cite{myerstalk}.

\begingroup\raggedright\endgroup

\end{document}